\documentstyle[preprint,aps]{revtex}

\begin{document}
\title{Inflation in a Flat Universe}
\author{K. Karaca\thanks{%
karacak@metu.edu.tr} , S. \c{S}. Bay\i n\thanks{%
bayin@newton.physics.metu.edu.tr}}
\address{{\it Department of Physics, Middle East Technical University}\\
{\it 06531, Ankara, Turkey}}
\date{March 01, 2002}
\maketitle

\begin{abstract}
We started the evolution of a flat universe from a nonsingular state called
prematter which is governed by an inflationary equation of state $P=(\gamma
-1)\rho ,$ where $\gamma $ represents the initial vacuum dominance of the
universe. The evolution of the universe except in the prematter era is
affected neither by the initial vacuum dominance nor by the initial
expansion rate of the universe. On the other hand, present properties of the
universe such as Hubble constant, age and density are sensitive to the value
of the temperature at the decoupling $\left( T_{m}\right) $. Over a range
between $3\cdot 10^{4}$and $5\cdot 10^{4}$ $K$ for $T_{m},$ our model
predicts a value between $50$ and $80$ $Km\cdot s^{-1}\cdot Mpc^{-1}$ for
the present value of the Hubble constant $\left( H_{0}\right) .$ Assuming
that the thermal history of the universe is independent from its geometry,
above range could be considered as a transition range for the decoupling
temperature $T_{m}.$
\end{abstract}

\section{Introduction}

The standard model of hot big-bang cosmology exhibits difficulties stemming
from the puzzling initial conditions. These difficulties are the well known
singularity, flatness, horizon, causality, homogeneity and isotropy
problems. During the past two decades, several models have been proposed to
overcome these difficulties. One common feature in all of these works is
that a new era called ``inflationary era'' was added into the history of the
universe before the radiation era with which the standard cosmological model
of the universe begins. During this new era, the universe expands enormously
from a finite size. This huge expansion is called ``inflation'' and arises
due to the unusual characteristic of the equation of state $(\ P=\left(
\gamma -1\right) \rho )$ used to describe the universe in this era.
Inflation mechanism was first suggested as a rescue from the horizon and
flatness problems [1] and then extended in such a way to construct models
which are free from an initial singularity. The equation of state used in
the inflationary era determines how much initially the universe is close to
the vacuum state and this is done by adjusting the values of $\gamma $.

Therefore, the models constructed in such a way, are parametric universe
models and evolution of these universes are also sensitive to the value of $%
\gamma $ in the inflationary era [2-5]. In other words, the present
properties of the universe such as the Hubble constant $\left( H\right) $
and the age $\left( t_{now}\right) $ are affected by the vacuum dominance of
the early universe. Besides the form of the equation of state, initial
expansion rate of the scale factor $\left( \dot{a}\right) $ also influences
the present properties of the universe and is considered as a parameter of
the universe model [5].

In the standard model, flatness problem arises from the extreme fine tuning
of the initial values of the energy density $\rho $ and Hubble constant $H$,
so that $\rho $ is very close to $\rho _{c}\left( \equiv 3H^{2}/8\pi \right) 
$ [1]$.$ This is necessary to produce a universe surviving $\sim 10^{10}$
years which is an age prediction compatible with the observational results.
In the models of inflationary cosmology, without a fine tuning requirement,
realistic universe models have been obtained. That is, flatness problem is
solved in these models. The solution to this problem could be attributed to
the enormous expansion of the universe during the inflationary era which
lasts a period at the order of the Planck scale $\sim 10^{-44}\sec .$ This
carries the geometry of the universe to the flat one while the value of $%
\Omega (\rho /\rho _{c})$ is driven toward one.

In this work, we construct a universe model having a flat space-time
geometry with the same motivations as in the other inflationary universe
models. Since the space-time geometry of this universe is already flat,
there is no flatness problem in this universe. On the other hand, the
universe is vacuum dominated due to the form of the equation of state used
to describe the universe in the inflationary era. This would determine how
long the universe would stay at the inflationary era before it enters the
radiation era predicted by the standard model. In the other inflationary
universe models, the present properties of the universe have been affected
by the initial vacuum dominance of the early universe. In this work, we will
study the effect of the initial vacuum dominance on the evolution of the
universe having flat space-time geometry. This would give us an idea about
the role of inflation mechanism when there is no flatness problem. This
would be interesting because inflation mechanism was first suggested as a
solution to the flatness problem [1] and the answer to the question: ``What
would be the response of a universe having a flat space-time geometry when
it is inflated through such a mechanism?'' is not exactly known. Our aim in
this paper would be to answer this question in a different context.

This paper is organized as follows: In the next section, we give the
dynamical equations of our model and solve them analytically. In Sec. III,
we give numerical results of the predictions of our model for the present
properties of the universe. We present our discussions and conclusions in
the final section.

\section{Dynamics of the Model}

\subsection{Field Equations}

The space-time geometry of the model describing a spatially homogenous and
isotropic universe is determined by a RW line element:

\begin{equation}
ds^{2}=dt^{2}-a^{2}(t)\left[ \left. dr^{2}+r^{2}\left( d\theta ^{2}+\sin
^{2}\theta d\phi ^{2}\right) \right] \right.
\end{equation}
where $(t,r,\theta ,\phi )$ are comoving coordinates, $a(t)$ is the scale
factor which represents the size of the universe.

For this line element, Einstein's gravitational field equations:

\begin{equation}
R_{\mu \nu }-%
{\displaystyle{1 \over 2}}%
g_{\mu \nu }R=-8\pi T_{\mu \nu },
\end{equation}
give

\begin{equation}
2\frac{\ddot{a}}{a}+\left( \frac{\dot{a}}{a}\right) ^{2}=-8\pi P,
\end{equation}

\begin{equation}
\left( \frac{\dot{a}}{a}\right) ^{2}=\frac{8\pi }{3}\rho ,
\end{equation}
where $P$ and $\rho $ are the energy density and pressure respectively in
the universe filled with a cosmological fluid well approximated by a perfect
fluid and a dot denotes differentiation with\ respect to the \ cosmic time $%
t $.

Eqs. (3) and (4) could be written as a single equation as

\begin{equation}
{\displaystyle{\ddot{a} \over a}}%
+4\pi \left( \gamma -\frac{2}{3}\right) \rho =0,
\end{equation}
where we have made use of the general form of the equation of state given as 
$P=\left( \gamma -1\right) \rho ,$ and $\gamma $ is assumed to be a constant
parameter during each era in the history of the universe.

From Eqs. (4) and (5), we could obtain an equation involving only the scale
factor $a(t)$ as

\begin{equation}
\left( 
{\displaystyle{\ddot{a} \over a}}%
\right) +\left( 
{\displaystyle{3 \over 2}}%
\gamma -1\right) \left( 
{\displaystyle{\dot{a} \over a}}%
\right) ^{2}=0.
\end{equation}
This equation could be solved for any $\gamma $ if we define conformal time $%
\eta $ as

\begin{equation}
dt=a\left( \eta \right) d\eta .
\end{equation}
In conformal time $\eta ,$ Eqs. (4), (5) and (6) become

\begin{equation}
\left( \frac{a^{\prime }}{a^{2}}\right) ^{2}=\frac{8\pi }{3}\rho ,
\end{equation}

\begin{equation}
\frac{a^{\prime \prime }}{a^{3}}-\left( \frac{a^{\prime }}{a^{2}}\right)
^{2}+4\pi \left( \gamma -\frac{2}{3}\right) \rho =0,
\end{equation}

\begin{equation}
\frac{a^{\prime \prime }}{a}+\left( \frac{3}{2}\gamma -2\right) \left( \frac{%
a^{\prime }}{a}\right) ^{2}=0,
\end{equation}
where a prime denotes differentiation with respect to the conformal time. If
we make the following change of variable

\begin{equation}
u\equiv 
{\displaystyle{a^{\prime } \over a}}%
=%
{\displaystyle{d\ln a \over d\eta }}%
,
\end{equation}
Eq. (10) could be brought into a much simpler form given as

\begin{equation}
u^{\prime }+cu^{2}=0,
\end{equation}
where

\begin{equation}
c=%
{\displaystyle{3 \over 2}}%
\gamma -1.
\end{equation}

Eq. (12) is a kind of Riccati equation and since we will consider vacuum
like state for the inflationary era $(c\neq 0),$ it could be solved by
setting

\begin{equation}
u=%
{\displaystyle{1 \over c}}%
{\displaystyle{w^{\prime } \over w}}%
=\left[ \ln \left( w^{1/c}\right) \right] ^{\prime }.
\end{equation}
This new change of variable leads us to write 
\begin{equation}
\frac{w^{\prime \prime }}{w}=0,
\end{equation}
which gives the solution in the following form

\begin{equation}
a\left( \eta \right) =a_{0}\left( \eta +b\right) ^{1/c},
\end{equation}
where $a_{0}$ and $b$ are integration constants.

Due to the unusual characteristic of the equation of state used in the
inflationary era, the temperature of the universe rises in this era although
the expansion of the universe is assumed to be adiabatic. This could be seen
if one considers the following relation derived from the first law of
thermodynamics:

\begin{equation}
{\displaystyle{T^{\prime } \over T}}%
+3%
{\displaystyle{a^{\prime } \over a}}%
\left( \gamma -1\right) =0
\end{equation}
(corresponding to Eq. (33) in [5]). Our modelling of the universe lies on a
thermodynamic assumption that the expansion of the universe continues until
the Planck temperature $T_{pl}$ which is assumed to be the maximum allowed
temperature. Then the equation of state takes the form of that of the
isotropic radiation. In other words, the universe starts behaving as
predicted in the standard model of the universe. Therefore, we give an
outline of our toy model as

a) {\bf The inflationary (prematter) era }$\left( 0\leq \eta \leq \eta
_{r}\right) :$ The equation of state is given as $P=\left( \gamma -1\right)
\rho $ where $\gamma =\gamma _{p}\simeq 10^{-3}.$ The constitution and
behavior of the material substance filling the universe in this era is not
exactly known.

b) {\bf The radiation era }$\eta _{r}\leq \eta \leq \eta _{m}):${\bf \ }The
universe is assumed to be composed of isotropic radiation for which the
equation of state is known to be $P=\rho /3.$ This corresponds to $\gamma
=\gamma _{r}=4/3$ in the general form of the equation of state.

c) {\bf The matter era }$\left( \eta _{m}\leq \eta \right) :$ The era that
we live in. Due to the large intergalactic distance and small relative
motions of the intergalactic objects, it would be assumed that the universe
is assumed to be filled with zero pressure dust such that $P=0,$ $\gamma
=\gamma _{m}=1.$

These eras are connected by first order phase transitions occurring at some
specific temperatures $T_{pl}$ and $T_{m}.$ $T_{m}$ is the temperature at
which radiation and matter are decoupled. There is not a generally accepted
value for this temperature other than ones which are speculated [6,7]. Due
to the lack of information, we will not assign a numerical value to $T_{m}$
until we find expressions about the present properties of the universe.

\subsection{Boundary conditions and solutions for the scale factor}

To eliminate the initial singularity in our model, we start the evolution of
the universe from a limiting density called the Planck density $\left( \rho
_{pl}\right) $\ corresponding to finite size for the scale factor of the
universe. To solve for the dynamics of the universe, we then specify the
initial expansion rate of the universe as 
\begin{equation}
a^{\prime }\left( 0\right) =v,
\end{equation}
where $v$ is a positive constant.

Writing\ Eq. (8) at $\eta =0,$ we get

\begin{equation}
a(0)=\sqrt{\frac{v}{\sqrt{d}}},
\end{equation}
which reflects the singularity-free character of our cosmological model. The
solutions for the scale factor in different eras are\ 
\begin{equation}
a(\eta )=\left\{ 
\begin{array}{ll}
a_{0}^{(p)}(\eta +b_{p})^{1/c_{p}}\  & \ \ 0\leq \eta \leq \eta _{r}, \\ 
a_{0}^{(r)}(\eta +b_{r})] & \eta _{r}\leq \eta \leq \eta _{m}, \\ 
a_{0}^{(m)}(\eta +b_{m})^{2}\  & \ \eta _{m}\leq \eta .
\end{array}
\right.
\end{equation}

We next impose that the scale factor and its derivative are continuous at
points $(\eta _{r},\eta _{m})$ where the phase transitions take place. This
leads us to determine the integration constants:

\begin{equation}
a_{0}^{(p)}=%
{\displaystyle{\sqrt{%
{\displaystyle{v \over H(0)}}} \over b_{p}^{\frac{1}{c_{p}}}}}%
,
\end{equation}

\begin{equation}
b_{p}=\frac{1}{c_{p}\sqrt{vH(0)}},
\end{equation}

\begin{equation}
a_{0}^{(r)}=\frac{a_{0}^{(p)}}{c_{p}}(\eta _{r}+b_{p})^{\frac{1}{c_{p}}-1},
\end{equation}

\begin{equation}
b_{r}=c_{p}b_{p}+(c_{p}-1)\eta _{r},
\end{equation}

\begin{equation}
a_{0}^{\left( m\right) }=\frac{a_{0}^{(r)}}{4(\eta _{m}+b_{r})},
\end{equation}

\begin{equation}
b_{m}=\eta _{m}+2b_{r,}
\end{equation}
where $H(0)$ is the initial value of the Hubble constant which is given as

\begin{equation}
H(0)=%
{\displaystyle{a^{\prime }(0) \over a^{2}(0)}}%
=\sqrt{d},
\end{equation}

\subsection{Physical properties of the model}

During the radiation era, the universe is assumed to consist of pure
radiation which is in thermal equilibrium. Therefore, its energy density
could be expressed in the blackbody form which is given as 
\begin{equation}
\rho _{blackbody}=%
{\displaystyle{8\pi ^{5}(kT_{pl})^{4} \over 15}}%
,
\end{equation}
where $k=1.38\cdot 10^{-16}K^{-1}$ is the Boltzmann constant.

As mentioned before, the universe heats up while expanding during the
inflationary era. We end the inflationary era with a thermodynamical
constraint which is the attainment of the Planck temperature $T_{pl}.$
Knowing that the scale factor of the universe evolves with the energy
density according to the Einstein's equations, we consider Eq. (8) together
with Eqs. (20), (21), (27) and (28), and end up with 
\begin{equation}
\eta _{r}=b_{p}\left[ \left( 1.5201\right) ^{\frac{c_{p}}{2\left(
c_{p}+1\right) }}-1\right] ,
\end{equation}
which is the conformal time corresponding to the phase transition between
the inflationary and radiation eras.

With the help of the Eq. (17), we could match the evolution of the scale
factor to that of the temperature. To this end, we have to specify $\gamma $
for each era. It is already mentioned that $\gamma $ equals $4/3$ for the
radiation era. Then we write 
\begin{equation}
{\displaystyle{a(\eta _{r}) \over a(\eta _{m)}}}%
=%
{\displaystyle{T_{m} \over T_{pl}}}%
,
\end{equation}
which, when combined with Eq. (24) gives 
\begin{equation}
\eta _{m}=c_{p}b_{p}\left( 1.5201\right) ^{\frac{c_{p}}{2\left(
c_{p}+1\right) }}\frac{T_{pl}}{T_{m}}-b_{r}
\end{equation}
which is the conformal time associated to the phase transition between
radiation and matter eras.

Before the emergence of matter as the dominant constituent of the energy
density in the universe, radiation decouples from matter and basic
constituents of matter (electrons, protons, neutrons etc.) start to form.
Then they combine\ for the first time in a process called ``recombination''
to form the more complex form of matter in the universe. At this point, it
is to be noted that after the decoupling between radiation and matter eras,
radiation still behaves as a perfect fluid responsible for the temperature
of the universe. As Kolb and Turner [6], we will consider decoupling as the
beginning of the matter era. With these in mind, we get from Eq. (17)

\begin{equation}
{\displaystyle{a\left( \eta _{m}\right)  \over a\left( \eta _{now}\right) }}%
=%
{\displaystyle{T_{now} \over T_{m}}}%
,
\end{equation}
where we have chosen integration limits to be the conformal times
corresponding to the second phase transition and the present time. This
equation could be solved for $\eta _{now}$ as

\begin{equation}
\eta _{now}=\sqrt{\frac{T_{m}}{T_{now}}}\eta _{m}+\left( \sqrt{\frac{T_{m}}{%
T_{now}}}-1\right) b_{m}
\end{equation}

In order to find comoving times corresponding to conformal times, we first
consider the definition given by Eq.(7) and assume that $t=0$ at $\eta =0$.
Then we get from Eq. (16)

\begin{equation}
t\left( \eta \right) =a_{0}\int_{0}^{\eta }\left( \eta ^{\prime }+b\right)
^{1/c}d\eta ^{\prime }.
\end{equation}
Since $c$ takes different values for each era in the history of the
universe, this integral has to be computed separately for each era. For the
inflationary $\left( c=c_{p}\right) ,$ radiation $\left( c_{r}=1\right) $
and matter $\left( c_{m}=1/2\right) $ eras, Eq. (34) can be integrated to
yield analytical expressions as

\begin{equation}
t_{r}=\frac{\sqrt{1.5201}-1}{\left( c_{p}+1\right) \sqrt{d}},
\end{equation}

\begin{equation}
t_{m}=t_{r}+\frac{\sqrt{1.5201}}{2\sqrt{d}}\left[ \left( \frac{T_{pl}}{T_{m}}%
\right) ^{2}-1\right] ,
\end{equation}

\begin{equation}
t_{now}=t_{m}+\frac{2}{3}\sqrt{\frac{1.5201}{d}}\left( \frac{T_{pl}}{T_{m}}%
\right) ^{2}\left[ \left( \frac{T_{m}}{T_{now}}\right) ^{3/2}-1\right] .
\end{equation}

We make use of the definition given as

\begin{equation}
H\left( \eta \right) \equiv \frac{a^{\prime }\left( \eta \right) }{%
a^{2}\left( \eta \right) },
\end{equation}
to find the Hubble constant at $\eta _{r},\eta _{m}$ and $\eta _{now}.$ They
are

\begin{equation}
H\left( \eta _{r}\right) =%
{\displaystyle{a^{\prime }\left( \eta _{r}\right)  \over a^{2}\left( \eta _{r}\right) }}%
=1.3436\cdot 10^{63}\ Km\cdot s^{-1}\cdot Mpc^{-1},
\end{equation}

\begin{equation}
H\left( \eta _{m}\right) =%
{\displaystyle{a^{\prime }\left( \eta _{m}\right)  \over a^{2}\left( \eta _{m}\right) }}%
=6.6925\cdot 10^{-2}\cdot T_{m}^{2}\text{ }Km\cdot s^{-1}\cdot Mpc^{-1},
\end{equation}

\begin{equation}
H\left( \eta _{now}\right) =%
{\displaystyle{a^{\prime }\left( \eta _{now}\right)  \over a^{2}\left( \eta _{now}\right) }}%
=0.2969\cdot \sqrt{T_{m}}\text{ }Km\cdot s^{-1}\cdot Mpc^{-1}.
\end{equation}

The energy density evolves with the scale factor of the universe as
described by Eq. (4) which when combined with Eq. (5) gives

\begin{equation}
\dot{\rho}+3\gamma 
{\displaystyle{\dot{a} \over a}}%
\rho =0.
\end{equation}
In terms of conformal time $\eta ,$ this equation becomes 
\begin{equation}
\rho ^{\prime }+3\gamma 
{\displaystyle{a^{\prime } \over a}}%
\rho =0,
\end{equation}
which could be solved as

\begin{equation}
{\displaystyle{\rho \left( \eta _{f}\right)  \over \rho \left( \eta _{i}\right) }}%
=\left( 
{\displaystyle{a\left( \eta _{i}\right)  \over a\left( \eta _{f}\right) }}%
\right) ^{3\gamma },
\end{equation}
where $\eta _{i}$ and $\eta _{f}$ mark the initial and final instants
respectively of any conformal time interval in a given era. When $\eta _{i}$
and $\eta _{f}$ \ are chosen as$\ \left( \eta _{i,}\eta _{f}\right) =\left(
0,\eta _{r}\right) ,\left( \eta _{r},\eta _{m}\right) ,\left( \eta _{m},\eta
_{now}\right) $ we obtain,respectively

\begin{equation}
\rho \left( \eta _{r}\right) =3.3923\cdot 10^{93}gr\cdot cm^{-3},
\end{equation}

\begin{equation}
\rho \left( \eta _{m}\right) =8.4166\cdot 10^{-36}\cdot T_{m}^{4}\text{ }%
gr\cdot cm^{-3},
\end{equation}

\begin{equation}
\rho \left( \eta _{now}\right) =1.6566\cdot 10^{-34}\cdot T_{m}\text{ }%
gr\cdot cm^{-3}.
\end{equation}

\subsection{Numerical Results}

The results that we have found for the physical properties of the universe
during its evolution clearly indicates that our model is sensitive to the
temperature at the last phase transition $\left( T_{m}\right) .$ On the
other hand, except the lifetime of the inflationary era, the time periods
are insensitive to the choice of parameter $\gamma _{p}.$ Since we do not
have enough information about what the value of $T_{m}$ should be, we will
try to assign some numerical results in the light of the recent
observational results for the present properties of the universe such as
Hubble constant, age and density. While doing this, as mentioned previously,
we keep in mind that $\gamma _{p}$ is no longer a parameter which affects
the evolution of the universe. Therefore, $T_{m}$ could be regarded as the
only parameter of our model.

Recent observations suggest a value approximately between $9$ and $15$ $Gyr$
for the present age of the universe $\left( t_{now}\right) $ [8,9] and a
value in the range of $\sim $ $50-80$ $Km\cdot s^{-1}\cdot Mpc^{-1}$ for the
present value of the Hubble constant $(H_{0})$ [10-16]. In our model, $T_{m}$
should take a value falling roughly into the range of $3\cdot 10^{4}-5\cdot
10^{4}$ $K$ so that the predictions of the model for $H_{0}$ and $t_{now}$
agree with observations. Since $\gamma _{p}$ has no effect on the evolution
of the universe, we fix it to $1.9000\cdot 10^{-3}$ and we present the
results about the present properties of the universe and those at the end of
each era for $T_{m}=3\cdot 10^{4}$ $K,$ $3.5\cdot 10^{4}$ $K,$ $4\cdot
10^{4} $ $K,$ $4.5\cdot 10^{4}$ $K,$ $5.0\cdot 10^{4}$ $K$ in tables 1-5.

\subsection{Discussion}

In this work, we try to explore the connection between flatness problem and
inflation mechanism. We do this by constructing a flat universe model by
adding a new era called inflationary era into the history of the universe.
In our model, the universe starts its journey with a vacuum like state and
undergoes a first order phase transition into the radiation era which is
predicted by the standard model as the era with which the universe first
starts to expand.

It is an already known fact that inflation mechanism carries different types
of space-time geometries (closed, open) to flat one by causing a huge
expansion in a time period at the order of Planck-Scale. Unlike other models
in which initial vacuum like structure of the universe affected the present
properties of the universe, in this model the radiation and matter eras are
not affected by the vacuum dominance of the early universe. The vacuum
dominance of the early universe is important when the universe is in the
inflationary era. If the universe starts with a pure vacuum state $(\gamma
=0)$, it cannot exit from inflation. This could be seen from the singular
behavior of $t_{r}$ for $\gamma =0.$ As the initial state of the universe
approaches the vacuum state, the time spent by the universe in the
inflationary era increases. After the universe exits from inflation, it
behaves as predicted in the standard model in which there is no initial
vacuum dominance.

The present properties of the universe such as Hubble constant, age and
density are not affected by the initial vacuum dominance. Instead, they are
highly sensitive to the temperature at the last phase transition $\left(
T_{m}\right) .$ In that case, this temperature can be considered as a
parameter. Therefore, the model we construct is a one-parameter universe
model. Considering the fact that the thermal history of the universe is
independent of the geometry of the universe, we may argue that the range
considered for $T_{m}$ could be seen as a transition range for the
decoupling temperature.

\bigskip

{\bf References}

[1] A. H. Guth, Phys. Rev. D, {\bf 23}, 347 (1981).

[2] M. Israelit, N.Rosen, Astrophys. J. {\bf 342}, 627 (1989) (IR).

[3] S.P. Starkovich, F. I. Cooperstock, Astrophys. J. {\bf 398}, 1 (1992)

\ \ \ \ \ (SC).

[4] S. S. Bayin, F. I. Cooperstock, V. Faraoni, Astrophys. J. {\bf 428}, 439
(1994)

\ \ \ \ \ (BCF).

[5] K. Karaca, S. \c{S}. Bayin (astro-ph/0007019).

[6] S. Weinberg, {\it Gravitation and Cosmology }(Wiley, New York, 1972).

[7] E. Kolb, M. Turner, {\it The Early Universe} (Addison-Wesley, New York,

\ \ \ \ \ 1990).

[8] B. Chaboyer et al., Astrophys. J. {\bf 494, }96 (1998).

[9] I. N. Reid, astro-ph/9704078.

[10] J. R. Mould et al., Astrophys. J. {\bf 529, }768 (2000).

[11] D. D. Kelson et al., Astrophys. J. {\bf 529, }768 (2000).

[12] L. Ferrarese et al., Astrophys. J. {\bf 529, }745 (2000).

[13] B. K. Gibson et al., Astrophys. J. {\bf 529, }723 (2000).

[14] S. Sakai et al., Astrophys. J. {\bf 529, }698 (2000).

[15] A. Saha et al., Astrophys. J. {\bf 522}, 802 (1999).

[16] R. Giovanelli et al., Astrophys. J. Lett. {\bf 477, }L1 (1997).

$\bigskip $ \newpage

\[
TABLE\text{ }1 
\]
\[
RESULTS\text{ }FOR\text{ }T_{m}=3\cdot 10^{4}K 
\]

\[
\begin{tabular}{|l|l|l|}
\hline\hline
$H(\frac{km}{s\cdot mpc})$ & $t(yr)$ & $\rho (\frac{gr}{cm^{3}})$ \\ 
\hline\hline
1.3436$\cdot 10^{63}$ & 4.8238$\cdot 10^{-50}$ & 3.3923$\cdot 10^{93}$ \\ 
\hline\hline
6.0233$\cdot 10^{7}$ & 8.1164$\cdot 10^{3}$ & 6.8174$\cdot 10^{-18}$ \\ 
\hline\hline
51.4246 & 1.2675$\cdot 10^{10}$ & 4.9698$\cdot 10^{-30}$ \\ \hline
\end{tabular}
\]

\[
TABLE\text{ }2 
\]
\[
RESULTS\text{ }FOR\text{ }T_{m}=3.5\cdot 10^{4}K 
\]

\[
\begin{tabular}{|l|l|l|}
\hline\hline
$H(\frac{km}{s\cdot mpc})$ & $t(yr)$ & $\rho (\frac{gr}{cm^{3}})$ \\ 
\hline\hline
1.3436$\cdot 10^{63}$ & 4.8238$\cdot 10^{-50}$ & 3.3923$\cdot 10^{93}$ \\ 
\hline\hline
8.1983$\cdot 10^{7}$ & 5.9631$\cdot 10^{3}$ & 1.2630$\cdot 10^{-17}$ \\ 
\hline\hline
55.5449 & 1.1735$\cdot 10^{10}$ & 5.7981$\cdot 10^{-30}$ \\ \hline
\end{tabular}
\]

\[
TABLE\text{ }3 
\]
\[
RESULTS\text{ }FOR\text{ }T_{m}=4\cdot 10^{4}K 
\]

\[
\begin{tabular}{|l|l|l|}
\hline\hline
$H(\frac{km}{s\cdot mpc})$ & $t(yr)$ & $\rho (\frac{gr}{cm^{3}})$ \\ 
\hline\hline
1.3436$\cdot 10^{63}$ & 4.8238$\cdot 10^{-50}$ & 3.3923$\cdot 10^{93}$ \\ 
\hline\hline
1.0708$\cdot 10^{8}$ & 4.5655$\cdot 10^{3}$ & 2.1546$\cdot 10^{-17}$ \\ 
\hline\hline
59.3800 & 1.0977$\cdot 10^{10}$ & 6.6264$\cdot 10^{-30}$ \\ \hline
\end{tabular}
\]

\[
TABLE\text{ }4 
\]
\[
RESULTS\text{ }FOR\text{ }T_{m}=4.5\cdot 10^{4}K 
\]

\[
\begin{tabular}{|l|l|l|}
\hline\hline
$H(\frac{km}{s\cdot mpc})$ & $t(yr)$ & $\rho (\frac{gr}{cm^{3}})$ \\ 
\hline\hline
1.3436$\cdot 10^{63}$ & 4.8238$\cdot 10^{-50}$ & 3.3923$\cdot 10^{93}$ \\ 
\hline\hline
1.3552$\cdot 10^{8}$ & 3.6073$\cdot 10^{3}$ & 3.4513$\cdot 10^{-17}$ \\ 
\hline\hline
62.9820 & 1.0349$\cdot 10^{10}$ & 7.4547$\cdot 10^{-30}$ \\ \hline
\end{tabular}
\]

\newpage

\[
TABLE\text{ }5 
\]
\[
RESULTS\text{ }FOR\text{ }T_{m}=5\cdot 10^{4}K 
\]

\[
\begin{tabular}{|l|l|l|}
\hline\hline
$H(\frac{km}{s\cdot mpc})$ & $t(yr)$ & $\rho (\frac{gr}{cm^{3}})$ \\ 
\hline\hline
1.3436$\cdot 10^{63}$ & 4.8238$\cdot 10^{-50}$ & 3.3923$\cdot 10^{93}$ \\ 
\hline\hline
1.6731$\cdot 10^{8}$ & 2.9219$\cdot 10^{3}$ & 5.2604$\cdot 10^{-17}$ \\ 
\hline\hline
66.3889 & 9.8178$\cdot 10^{9}$ & 8.2830$\cdot 10^{-30}$ \\ \hline
\end{tabular}
\]

\end{document}